\documentclass[utf8]{frontiersFPHY_without_logo} 

\setcitestyle{square} 
\usepackage{url,hyperref,lineno,microtype,subcaption}
\usepackage[onehalfspacing]{setspace}
\usepackage{bm}

\def\keyFont{\fontsize{8}{11}\helveticabold }
\def\firstAuthorLast{Fabian Jan Schwarzendahl \& Daniel Beller} 
\def\Authors{Fabian Jan Schwarzendahl\,$^{1,2,*}$, Daniel A. Beller\,$^{3,2,*}$}

\def\Address{$^{1}$Institut f\"ur Theoretische Physik II: Weiche Materie, Heinrich-Heine-Universit\"at D\"usseldorf, 40225 D\"usseldorf, Germany \\
$^{2}$Department of Physics, University of California, Merced, California 95343, USA \\
$^{3}$Department of Physics and Astronomy, Johns Hopkins University, Baltimore, Maryland 21218, USA}
\def\corrAuthor{Fabian Jan Schwarzendahl, Daniel A. Beller}

\def\corrEmail{fabian.schwarzendahl@hhu.de, d.a.beller@jhu.edu}

\begin{document}
\onecolumn
\firstpage{1}

\title[]{Do active nematic self-mixing dynamics help growing bacterial colonies to maintain local genetic diversity?} 

\author[\firstAuthorLast ]{\Authors} 
\address{} 
\correspondance{} 

\extraAuth{}

\maketitle


\begin{abstract}

Recent {studies} 
have shown that packings of cells, both eukaryotic cellular tissues and growing or swarming bacterial colonies, {can often be understood as active nematic fluids}. 
A key property of volume-conserving active nematic model systems is chaotic self-mixing characterized by motile topological defects. However, for active nematics driven by growth rather than motility, less is understood about mixing and defect motion.  Mixing could {affect evolutionary outcomes in} 
bacterial colonies by counteracting the tendency to spatially segregate into monoclonal sectors, which reduces the local genetic diversity {and confines competition between subpopulations to the boundaries between neighboring sectors}. To examine whether growth-driven active nematic physics could {influence this genetic demixing process,} 
we conduct agent-based simulations of growing, dividing, and sterically repelling rod-like bacteria of various aspect ratios, and we analyze colony morphology using tools from both soft matter physics and population genetics. We find that despite measurable defect self-propulsion in growth-driven active nematics, the radial expansion flow prevents chaotic mixing. Even so, at biologically relevant cell aspect ratios, self-mixing is more effective in growing active nematics of rod-like cells compared to growing isotropic colonies of round cells. This suggests {potential evolutionary consequences} 
associated with active nematic dynamics.

\tiny
 \keyFont{ \section{Keywords:} active matter, active nematics, microbial evolution, neutral evolution, genetic mixing, liquid crystals, topological defects, range expansion} 
\end{abstract}

\section{Introduction}
The study of active matter provides insights into the collective motions of living and internally driven systems by marrying far-from-equilibrium dynamics with the physics of partially ordered materials \cite{marchetti2013hydrodynamics,needleman2017active}. In co-moving groups of vertebrates such as flocks of birds and schools of fish, as well as in flocks of self-propelled colloids, the emergent order is typically polar orientational alignment analogous to ferromagnets \cite{marchetti2013hydrodynamics,bricard2013emergence,cavagna2014bird,menon2010active}. 
Meanwhile, active matter with the apolar orientational order of nematic liquid crystals occurs in vibrated granular rods and in a wide variety of systems at the microscale, including suspensions of cytoskeletal biofilaments and motor proteins \cite{decamp2015orientational,kumar2018tunable}, monolayers of eukaryotic cells \cite{duclos2014perfect,kawaguchi2017topological,saw2017topological}, and morphogenesis of the animal \textit{Hydra} \cite{maroudas2020topologicalPublished}. The non-equilibrium production of nematic disclinations is now recognized in a broad variety of 2D systems as a signature of active nematic physics  \cite{marchetti2013hydrodynamics,decamp2015orientational,giomi2013defect,giomi2014defect,shankar2018defect,duclos2020topological}. Activity produces pairs of $\pm 1/2$-winding number disclinations, with the $+1/2$ defect behaving like a self-propelled particle.

Swarms of motile bacteria are a paradigmatic example of active matter
\cite{wensink2012meso,bar2020self,schwarzendahl2018maximum,wioland2013confinement,wioland2016directed,copeland2009bacterial,dombrowski2004self,ishikawa2008coherent,copenhagen2021topological}. 
Recently, it has been shown that even non-motile bacteria can behave as an active nematic simply by growing, dividing, and pushing against their neighbors \cite{volfson2008biomechanical,sengupta2020microbial,you2018geometry,you2019mono,you2021confinement,doostmohammadi2016defect,dellarciprete2018growing,farrell2013mechanically,nordemann2020defect,farrell2017mechanical,shimaya20213d}. We will refer to this scenario as the ``growing active nematic'', in contrast to the better-studied active nematic of fixed size. 
Experiments have demonstrated that  \textit{E.\ coli}  colonies form nematic microdomains of well-aligned rods, with neighboring domains meeting at grain boundary lines punctuated by disclination point-defects of winding number $\pm 1/2$  \cite{dellarciprete2018growing,sengupta2020microbial,you2018geometry}. Agent-based simulations have found that simple mechanical models of growing, dividing, and mutually volume-excluding cells are sufficient to reproduce this nematic microdomain structure \cite{you2018geometry,nordemann2020defect}. For rod-like bacteria, these mechanical interactions create torques causing nearby cells to align their long axes in the apolar manner of nematic liquid crystals. 
 An observed steady-state density of disclinations, maintained by a balance of $\pm 1/2$ defect pair-annihilation with pair-creation, represents a key similarity between growing and fixed-size active nematic systems.   A similar non-equilibrium steady-state population of disclinations is found in continuum hydrodynamic simulations, which predict that $+1/2$ defects self-propel as they do in fixed-size active nematics \cite{doostmohammadi2016defect,dellarciprete2018growing,you2018geometry}. 

Chaotic self-mixing is one of the most striking features of fixed-size active nematics  \cite{decamp2015orientational,tan2019topological,duclos2020topological}, drawing comparisons to turbulent flow \cite{thampi2014vorticity,giomi2015geometry}. In the microtubule-kinesin active nematic {model system}, the chaotic  self-advection was quantified in Ref.~\cite{tan2019topological} through measures of material stretching including the Lyapunov exponent. 
{A strong correspondence was found between} 
the fluid self-mixing 
and the topological entropy measurement of braiding in the trajectories of $+1/2$ defects. This supports a coarse-grained picture of active nematic chaos in which the $+1/2$ defects are stirring rods, efficiently mixing the viscous fluid around them. 

Does defect-driven, chaotic self-mixing occur also in growing active nematics? 
The answer may impact the evolutionary fate of the colony because of an apparently opposite trend exhibited by growing colonies of immotile cells: Genetic demixing, or gene segregation, divides the expanding colony into sectors that each consist of only closely related cells  \cite{hallatschek2007genetic,korolev2010genetic,Banwarth2020Quantifying}. In radially growing colonies, many genetic sectors are lost to coarsening dynamics at early times, after which an ``inflationary'' epoch occurs in which sectors grow without interacting \cite{lavrentovich2013radial}.

This demixing is a particularly stark manifestation of the general loss of local genetic diversity found in more recently colonized areas when a species expands its range---whether arising from microbial colony growth in a Petri dish or from invasive species at scales up to the continental \cite{edmonds2004mutations,klopfstein2006fate}. Genetic drift, the random fluctuations in allele frequencies from one generation to the next, acts far more strongly at the expansion periphery than in the homeland, as new regions become dominated by the alleles that happened to be carried there first (founder effect) \cite{mayr1999systematics,hallatschek2007genetic}. Consequently, Darwinian selection has {comparatively} less influence with which to promote adaptation  \cite{elena2003evolution}, {leaving the population more vulnerable to \textit{e.g.} deleterious mutations, diseases, and environment changes.}

Taken together, genetic demixing and active self-mixing raise the possibility of evolutionary {consequences} 
for growing bacterial colonies exhibiting active nematic behavior: If growing active nematics experience the efficient physical mixing observed in fixed-size active nematics, then this internal reorganization may partially counteract or delay  genetic demixing. 


In considering the interplay between active nematic mixing and genetic demixing, we note that the two phenomena have mostly been studied under different nutrient conditions. For growing active nematics, researchers have so far assumed nutrient-rich conditions, meaning that bacterial growth rates are independent of the cell's position in the colony. In contrast, studies of genetic demixing usually assume nutrient availability to be low enough that growth occurs only within a short distance from the expansion front \cite{farrell2013mechanically, farrell2017mechanical}. However, gene segregation still occurs in the nutrient-rich (bulk-driven growth) scenario studied for active nematics, albeit with less sharp boundaries between genetic sectors \cite{Banwarth2020Quantifying}, as we will see below in small aspect-ratio rods (Fig.~\ref{fig:mixing_obs}a).

In this article, we computationally investigate whether defect-driven active self-mixing helps growing bacterial colonies to maintain local genetic diversity. Using a minimal mechanical model of growing and dividing rod-like cells, we study active nematic dynamics and population evolution in tandem, including a quantitative study of defect motion and gene segregation for cells of various aspect ratios. Our findings shed light on unexpected subtleties in comparing growing to non-growing active nematics, and on potential evolutionary {consequences} 
related to emergent active matter dynamics. 





\section{Methods}

\subsection{Computational model} 

\begin{figure}
    \centering
    \includegraphics[width=0.3\textwidth]{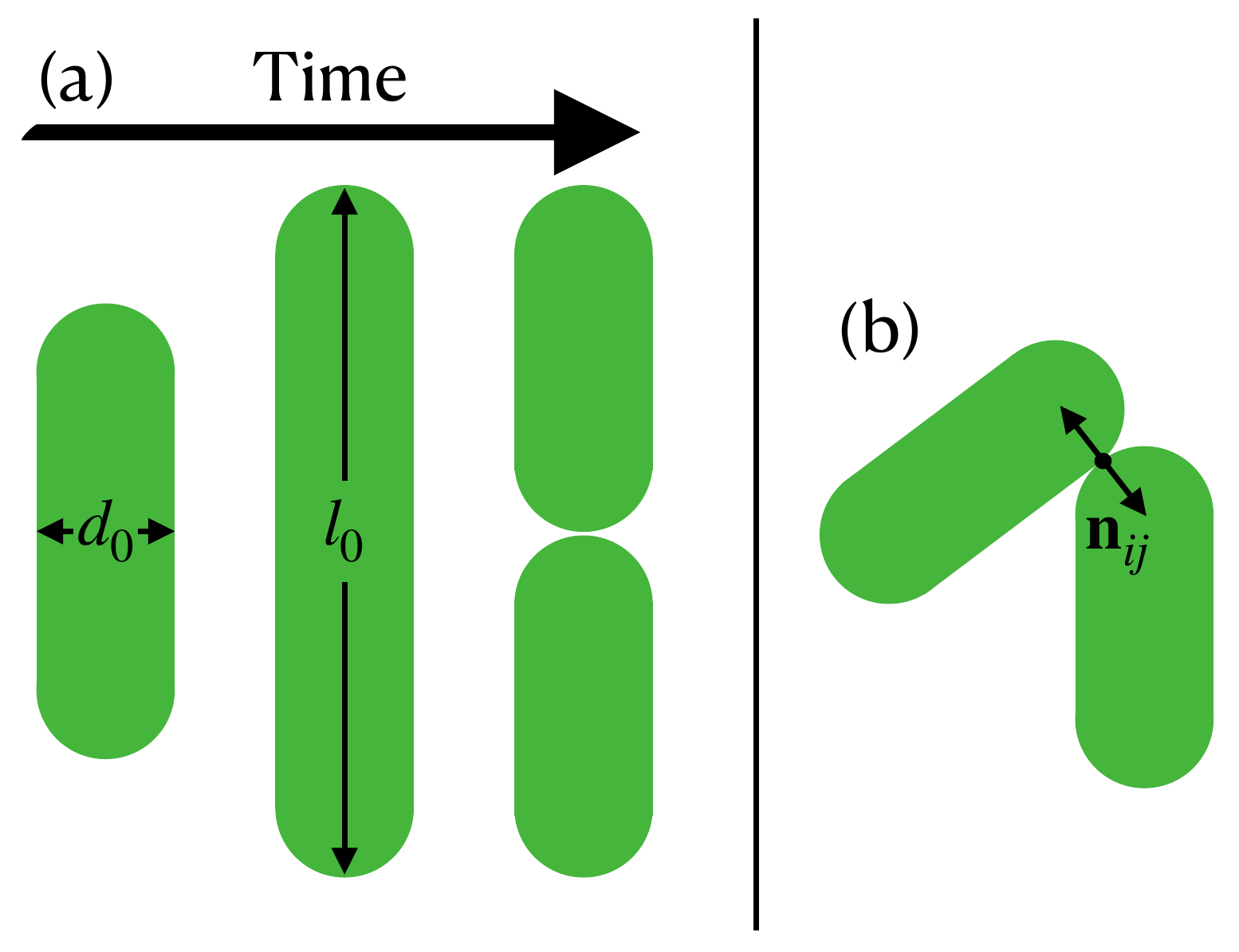}
    \caption{Growing bacteria modeled as expanding rods. (a) A rod grows, reaches  length $l_0$ and splits into two. (b) Interaction between two rods, where the normal vector $\bm{n}_{ij}$ of the particles' closest point of contact  is shown.}
    \label{fig:sketch}
\end{figure}

We model growing bacteria as two-dimensional expanding rods, whose diameter $d_0$ is fixed and whose length $l_i$ expands with a mean growth rate $g$ up to length $l_0$ (Fig.~\ref{fig:sketch}a).
When a rod reaches length $l_0$ it splits into two rods {of length $l_0/2$}. 
The growth of each rod with rate $g_i$ is linear in time, and for each rod we choose $g_i$ to be a random variable in the interval $[g/2,3g/2]$ to avoid synchronization effects \cite{you2018geometry}. 
Following mechanics-focused approaches used in recent computational studies of bacterial colonies \cite{rudge2012computational,you2018geometry,nordemann2020defect,smith2017cell,ghosh2015mechanically},  the rods' dynamics are governed by an overdamped motion for the positions $\bm{r}_i$ and orientations $\bm{e}_i=(\mathrm{cos}\theta_i, \mathrm{sin}\theta_i )^T$ expressed by the angle $\theta_i$. Explicitly, they are updated according to
\begin{align}
  &  \frac{\mathrm{d}
  \bm{r}_{i}}{\mathrm{d} t}=\frac{1}{\zeta l_{i}} \sum_{j} \bm{F}_{i j}, \\
&\frac{\mathrm{d} \theta_{i}}{\mathrm{d} t}=\frac{12}{\zeta l_{i}^{3}} \sum_{j}\left(\bm{r}_{i j} \times \bm{F}_{i j}\right) \cdot \bm{e}_z,
\end{align}
where $\zeta$ is a friction coefficient, $\bm{e}_z$ is the unit vector that is perpendicular to the plane of motion of the rods, $\bm{r}_{ij}= \bm{r}_i-\bm{r}_j$ is  the {separation} 
vector between two rods, and $\bm{F}_{ij}$ are the forces between rods.
The forces are modeled using a Hertzian repulsion, which reads 
\begin{align}
    \bm{F}_{i j}=F_0 d_{0}^{1 / 2} h_{i j}^{3 / 2} \bm{n}_{i j},
\end{align}
where $F_0$ is the repulsion strength, $h_{i j}$ is the overlap {length} of particle $i$ and $j$, and $\bm{n}_{ij}$ is the normal vector of the particles' closest point of contact (Fig.~\ref{fig:sketch}b).
In the following we use $l_0$ as a unit of length, and $g$ as a unit of velocity, implying a natural unit of time $l_0/g$. Our model has two dimensionless parameters, which are the aspect ratio $l_0/d_0$ and $g \zeta/(F_0 l_0)$; the latter we keep fixed to $g \zeta/(F_0 l_0)= 5 \times 10^{-7}$. We simulate up to $25000$ rods {from an initial state of a single rod}.

\subsection{Mapping to continuum fields} 

To analyze the simulated colonies, we compute a nematic {tensorial orientation} 
field from our particle-based simulations. Following Ref.~\cite{dellarciprete2018growing}, we {apply a smoothing function that transforms the $i$th particle into} 
\begin{align}
    h_i(\bm{r}) = 
    \frac{1}{4}\left( 
    \mathrm{tanh} \left( \frac{\bm{e}_i \cdot \Delta \bm{r}_i + l_i/2}{\sigma} \right) 
    -\mathrm{tanh} \left( \frac{\bm{e}_i \cdot \Delta \bm{r}_i - l_i/2}{\sigma} \right) 
    \right)
  \nonumber \\ \times
  \left( 
    \mathrm{tanh} \left( \frac{\bm{e}^\perp_i \cdot \Delta \bm{r}_i + d_0/2}{\sigma} \right) 
    -\mathrm{tanh} \left( \frac{\bm{e}^\perp_i \cdot \Delta \bm{r}_i - d_0/2}{\sigma} \right) 
    \right),
    \label{eq:smoothing}
\end{align}
{where $\Delta \bm{r}_i = \bm{r}-\bm{r}_i$ is the separation from the particle's position $\bm{r}_i$; $\bm{e}^\perp_i$ is the vector perpendicular to the particle's orientation $\bm{e}_i$; $l_i$ is the particle's current length;} and $\sigma=2 d_0$ is the smoothing length. 
The nematic field is then given by
\begin{align}
    Q_{\mu \nu}(\bm{r}) = \sum_i^N h_i(\bm{r}) \left(  2 e_{i,\mu} e_{i,\nu} - \delta_{\mu \nu}   \right),
\end{align}
where the sum runs over particle indices $i$ and $\mu,\nu$ refer to the vector components. Similarly, we calculate a velocity field as 
\begin{align}
    \bm{v}(\bm{r}) = \sum_i^N h_i(\bm{r}) \bm{v}_i,
    \label{eq:velfield}
\end{align}
where $\bm{v}_i= \mathrm{d} \bm{r}_i / \mathrm{d}t$ is the velocity of an individual particle approximated from its displacement in the most recent timestep.

\section{Results}

\begin{figure}
    \centering
    \includegraphics[width=1.0\textwidth]{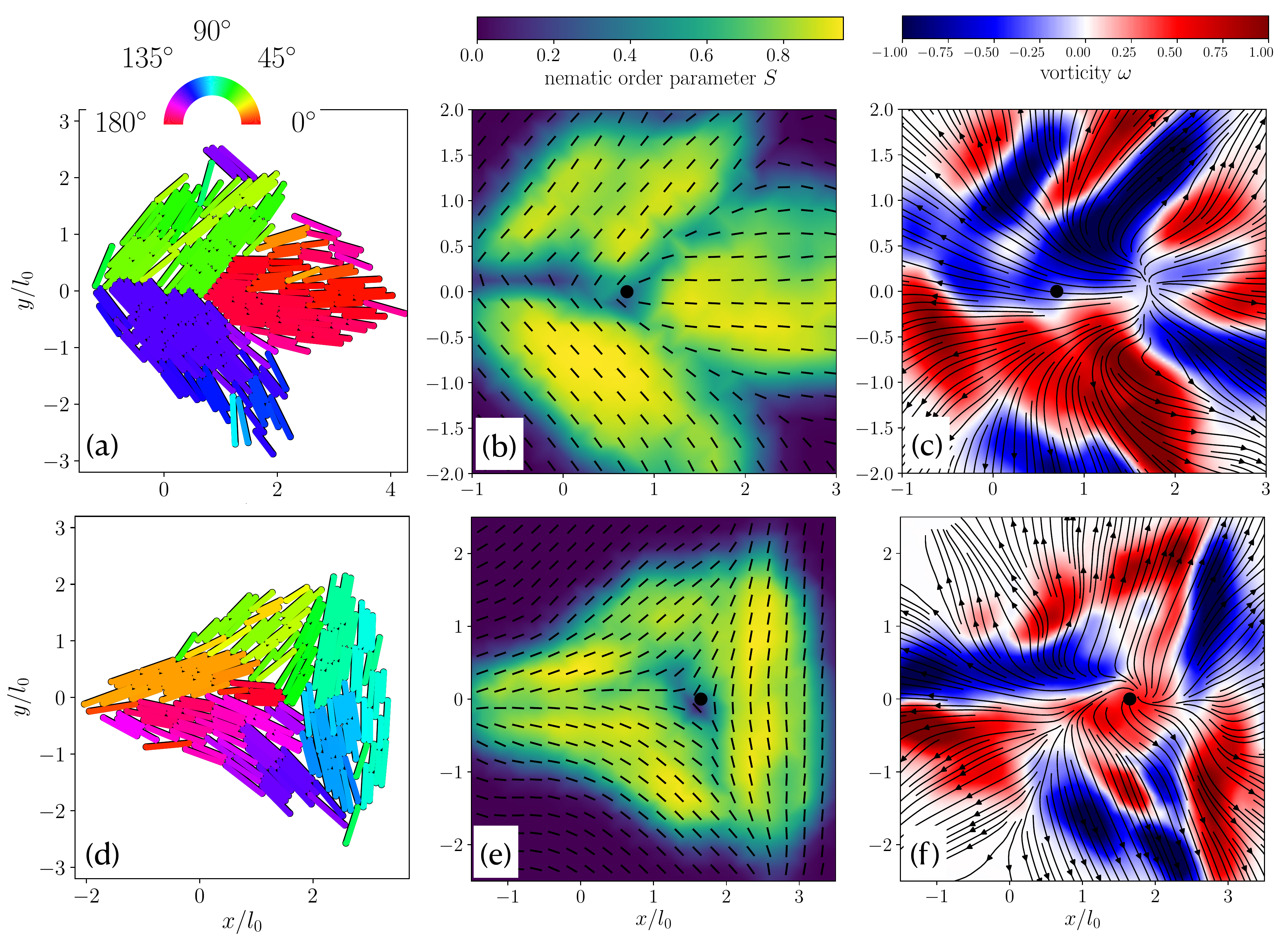}
    \caption{(a) Simulated configuration of an isolated $+1/2$ defect. Color code shows the orientation of each rod. (b) Nematic director field (black lines) and nematic order  parameter (color code) extracted from the configuration in (a). (c) Flow field (black  lines) and vorticity (color code) extracted from the configuration in (a). (d) Configuration of an isolated $-1/2$ defect. (e) Nematic director (black lines) and order parameter (color code) extracted from the configuration shown in (d). (f) Flow field (black lines) and vorticity (color code) extracted from (d). The black filled circle in (b),(c),(e) and (f) shows the location of the defect core. Particles have aspect ratio $l_0/d_0=10$.}
    \label{fig:defect_flows}
\end{figure}

\subsection{Velocity field for isolated nematic defects}

Before studying colonies containing many defects, we examine the early-time behavior of isolated nematic defects programmed into the colony's initial condition. In hydrodynamic field theories of fixed-size active nematics, director fields containing isolated defects display very distinct vorticity patterns depending on whether the defect has positive or negative charge: a $+1/2$ defect generates two regions of opposite vorticity nearby, resulting in a net active force in the transverse direction, whereas a $-1/2$ defect is surrounded by a threefold-symmetric pattern of alternating vorticity that produces no net force at the defect core~\cite{giomi2015geometry,giomi2014defect}. 

To investigate whether this picture applies in the case of growing active nematics,   we initialize our simulation with rod orientations approximating {an individual defect}. 
 We then {allow the colony to grow until each cell has  doubled at least once}, resulting in the configurations shown in Fig.~\ref{fig:defect_flows}a,d for a $-1/2$ and $+1/2$ defect respectively.
Figures~\ref{fig:defect_flows}b,e display the corresponding nematic field, exhibiting the typical structures of $+1/2$ and $-1/2$ defects. 

The velocity field reveals a flow generally oriented radially outward from the center of the colony for both $+1/2$ and $-1/2$ defects (see Fig.~\ref{fig:defect_flows}c,f), which is induced by the radial outward growth of the colony. We also computed the flow fields' vorticity $\omega= \frac{\partial v_y}{\partial x} - \frac{\partial v_x}{\partial y}$ (color code in Fig.~\ref{fig:defect_flows}c,f). In line with nematic field theories~\cite{giomi2014defect,brezin2022spontaneous} the $+1/2$ defect displays two regions of vorticity with opposite sign and the $-1/2$ defect shows a combination of six regions with alternating sign. These results for isolated defects lead us to expect that the $+1/2$ defect self-propulsion dynamics familiar from fixed-size active nematics will compete with the radial expansion flow in growing active nematics.


\begin{figure}
    \centering
    \includegraphics[width=1.0\textwidth]{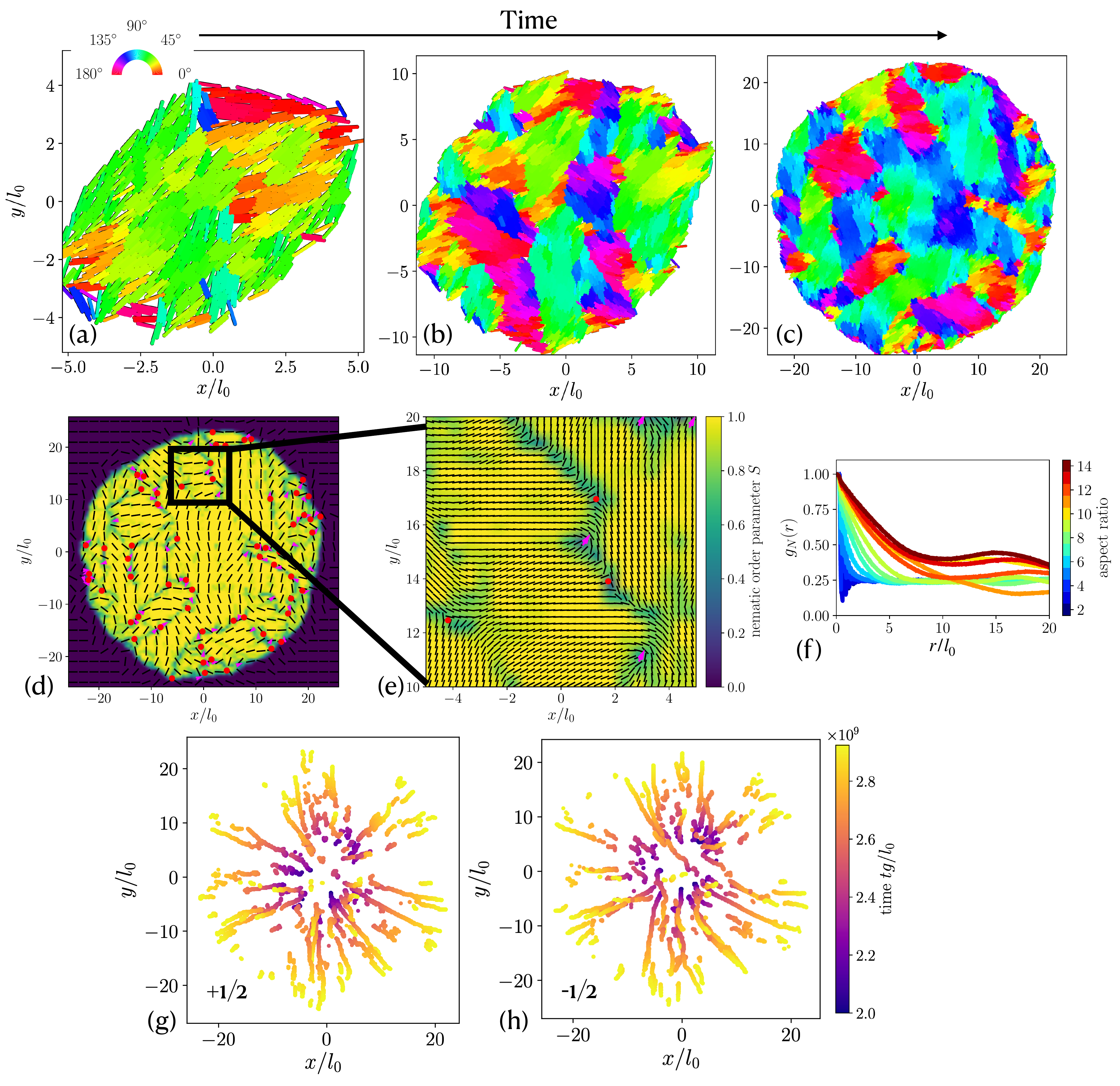}
    \caption{ (a)-(c) Snapshots of a simulated colony, {with particles colored by orientation}, at times {$tg/l_0 $ equal to (a) $2\times 10^9 $, (b) $2.5\times 10^9$, (c) $3\times 10^9$}.
     Particles have aspect ratio $l_0/d_0 = 10$. 
    (d) Nematic orientation {field} (black lines) and nematic order parameter 
    (color code) {calculated for} 
    the simulation snapshot shown in (c). Defects of charge $-1/2$ and $+1/2$ are marked with red dots and magenta arrows, respectively.
    (e) Zoom-in of the black box shown in (d).
    (f) Orientational correlation function $g_N(r)$ as a function of interparticle distance $r$, for colonies with various cell aspect ratios.
    (g)-(h) Trajectories of $+1/2$ (g) and $-1/2$ (h) defects extracted from the nematic field of the colony shown in (a)-(c).
   }
    \label{fig:snapshots}
\end{figure}

\subsection{{Nematic order and defects in a growing colony}}

Typical simulation snapshots {resulting from a single-cell initial condition} are shown in Fig.~\ref{fig:snapshots}a-c, displaying the growth of a colony. Consistent with prior work~\cite{you2018geometry}, we find microdomains of rods with similar orientation. {Neighboring microdomains meet at grain boundaries, many of which contain nematic point-disclination defects.} 
Figure~\ref{fig:snapshots}d shows a typical nematic field for a {colony at the end of a simulation,} 
where we {mark} 
nematic defects with $+1/2$ ({magenta} 
arrows)  and $-1/2$ (red circles) charge. The defects are identified by integrating {changes in orientation on a small circuit} around every point in the nematic field, leading to a number of defect candidates around {each} 
(true) defect position, which is then found using a clustering algorithm. 

As shown in Fig.~\ref{fig:snapshots}e, this method identifies defects reliably such that we can extract defect trajectories.  These are shown in Fig.~\ref{fig:snapshots}g for $+1/2$ defects and in Fig.~\ref{fig:snapshots}h for $-1/2$ {defects}. For both {defect types,} 
we observe a {predominant} 
radial outward trend, implying that the expansion of the colony has a strong influence on the defect motion. 

The absence of significant braiding among the $+1/2$ defect trajectories (Fig.~\ref{fig:snapshots}g) represents a notable distinction from fixed-size active nematics, where such braiding corresponds strongly to the active material's chaotic flow field \cite{tan2019topological}.
Relatedly, we find that Lyapunov exponents calculated for our system, using the velocity field of Eq.~\eqref{eq:velfield}, are statistically indistinguishable from zero, regardless of particle aspect ratio.

{As a measure of nematic order, we compute the orientational correlation function, commonly used for characterizing nematic liquid crystals~\cite{greschek2011finite}, which reads
\begin{align}
    g_N(r) = \langle P_2( \bm{e}_i \cdot \bm{e}_j) \rangle,
\end{align}
where $P_2(x)= \frac{1}{2} ( 3 x^2 -1)$ is the second Legendre polynomial and $r= |\bm{r}_i-\bm{r}_j|$ is the distance between particle $i$ and $j$. As we would expect, $g_N(r)$ decays slowly for large-aspect ratio particles, indicating nematic orientational correlations over some significant correlation length, whereas for smaller aspect ratios, a sharp decay of $g_N$ near $r=0$ reveals an absence of nematic order (Fig~\ref{fig:snapshots}f).
}

\begin{figure}
    \centering
    \includegraphics[width=1.0\textwidth]{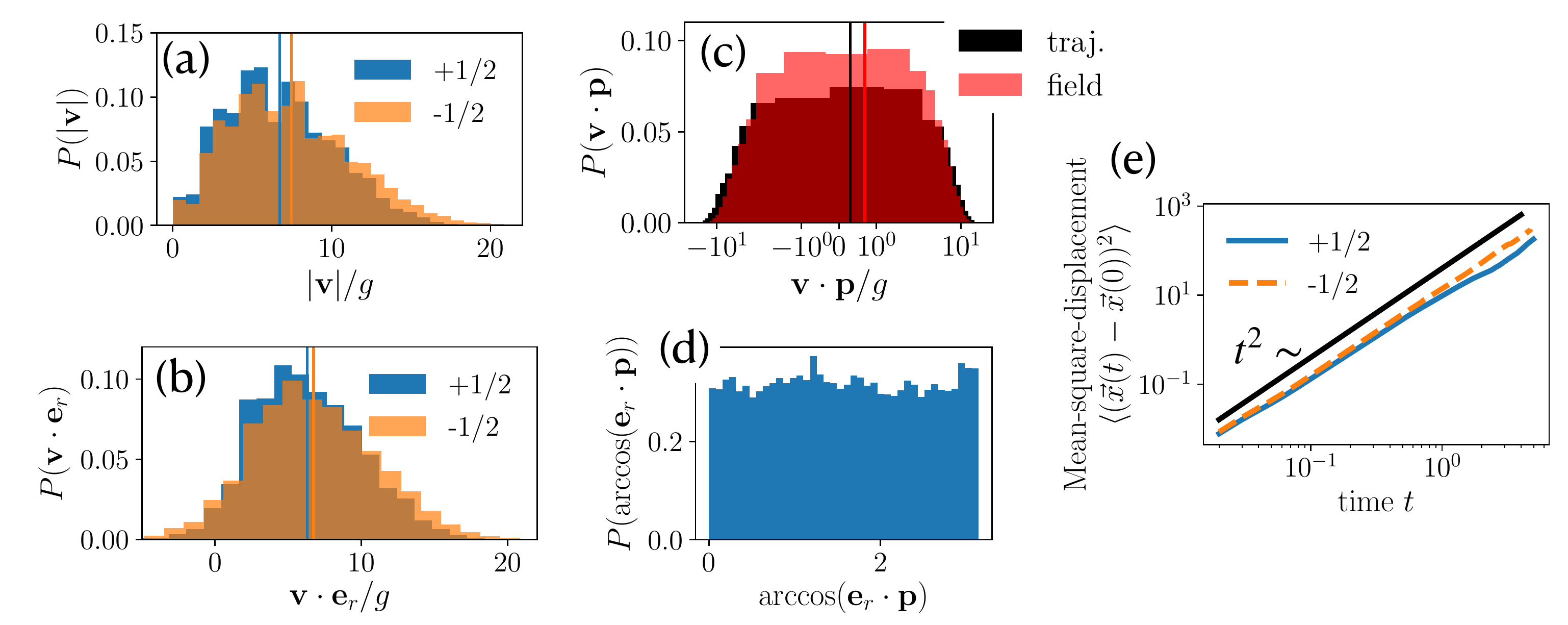}
    \caption{(a,b) Distributions of {speeds (a) and radial component of velocity (b)} 
    of $+1/2$ (blue) and $-1/2$ (orange) defects. Vertical lines show  mean of each distribution. 
    (c) Distribution of {velocity projected onto heading direction} 
    of $+1/2$ defects. For black bars the velocity was extracted from the defect trajectories while for the red bars the velocity was extracted from the flow field. Lines show the mean of the respective distributions. (d) Distribution of heading direction of $+1/2$ defects with respect to the radially outward unit vector from the colony's center of mass. (e) Mean-square displacements of $+1/2$ (blue solid line) and $-1/2$ (orange dashed line) defects. Black line  shows a $t^2$ scaling. We used an aspect ratio of $l_0/d_0=10$ and averaged over defect trajectories from 10 independent simulation runs.} 
    \label{fig:defect_dists}
\end{figure}

 \subsection{Motion of nematic defects}
We now examine the  statistical properties of the nematic defects' motion. First, we compute their velocities $\bm{v}$ and their {speed} 
distribution, shown in Fig.~\ref{fig:defect_dists}a. We find that both types of defects have a peak at intermediate {speeds,} 
which is more prominent for $+1/2$ defects. {Conversely, and surprisingly, higher speeds are somewhat more common for $-1/2$ defects.} 
A similar picture arises when we project the defects' {velocities} 
onto the radially (from the center of mass of the colony) outward {direction} 
at the defect's position (Fig.~\ref{fig:defect_dists}b). Again, both $+1/2$ and $-1/2$ defects show a peak {at a positive value}, and the $-1/2$ defects' distribution  is enhanced towards high {radial} velocities. That is, while colony growth generally pushes all defects farther from the colony center, outward radial motion is greater for $-1/2$ defects than for $+1/2$ defects. 




However, for $+1/2$ defects, we have a second contribution to their motion: the asymmetry in the nematic field around the defect leads to an active motion {or ``self-propulsion''}, a well-known phenomenon in {fixed-size active nematics}
~\cite{giomi2013defect,giomi2014defect,decamp2015orientational}.  {To determine whether, and how strongly, $+1/2$ defects self-propel in growing active nematics,} 
we compute {each defect's} 
heading direction $\bm{p}$, 
the {orientation} 
of the 
``comet-head'' {arrangement in the nearby director field, which is} 
computed using the polarization $p_{\mu}= -\partial_\nu Q_{\mu \nu}/|\partial_\nu Q_{\mu \nu}|$ at the defect core. We then {examine the projection of the defect's velocity onto its heading direction.} 
The defect's velocity is computed with two  independent methods, first from its trajectory  and second {from the velocity field $\bm{v}(\bm{r})$ at the defect's location.} 
The resulting distributions, {shown in Fig.~\ref{fig:defect_dists}c, both display a broad range of positive and negative values, but with a bias toward positive values. This implies that $+1/2$ defects move preferentially along their heading direction, the same defect self-propulsion trend found in fixed-size, extensile active nematics}. 
{We additionally find that the defect heading directions are isotropically distributed in the lab frame and have no significant correlation with the radial direction (Fig.~\ref{fig:defect_dists}d).}

The mean-square-displacement of defects (Fig.~\ref{fig:defect_dists}e) shows  a quadratic scaling {with time,} signifying ballistic motion, that stems from the radial expansion of the colony. Ballistic motion implies that the radial expansion dominates over the diffusive defect stirring dynamics typical of fixed-size active nematics \cite{giomi2015geometry,tan2019topological}.

From the distributions in Fig.~\ref{fig:defect_dists}, {along with the defect trajectories in Fig.~\ref{fig:snapshots}g,h,} the following conclusion {emerges:} 
{While $+1/2$ defect self-propulsion is detectable in growing active nematics,  the motion of $+1/2$ defects (as well as that of $-1/2$ defects) is dominated by the radial expansion flow.} 

\subsection{Genetic mixing}
\begin{figure}
    \centering
    \includegraphics[width=1.0\textwidth]{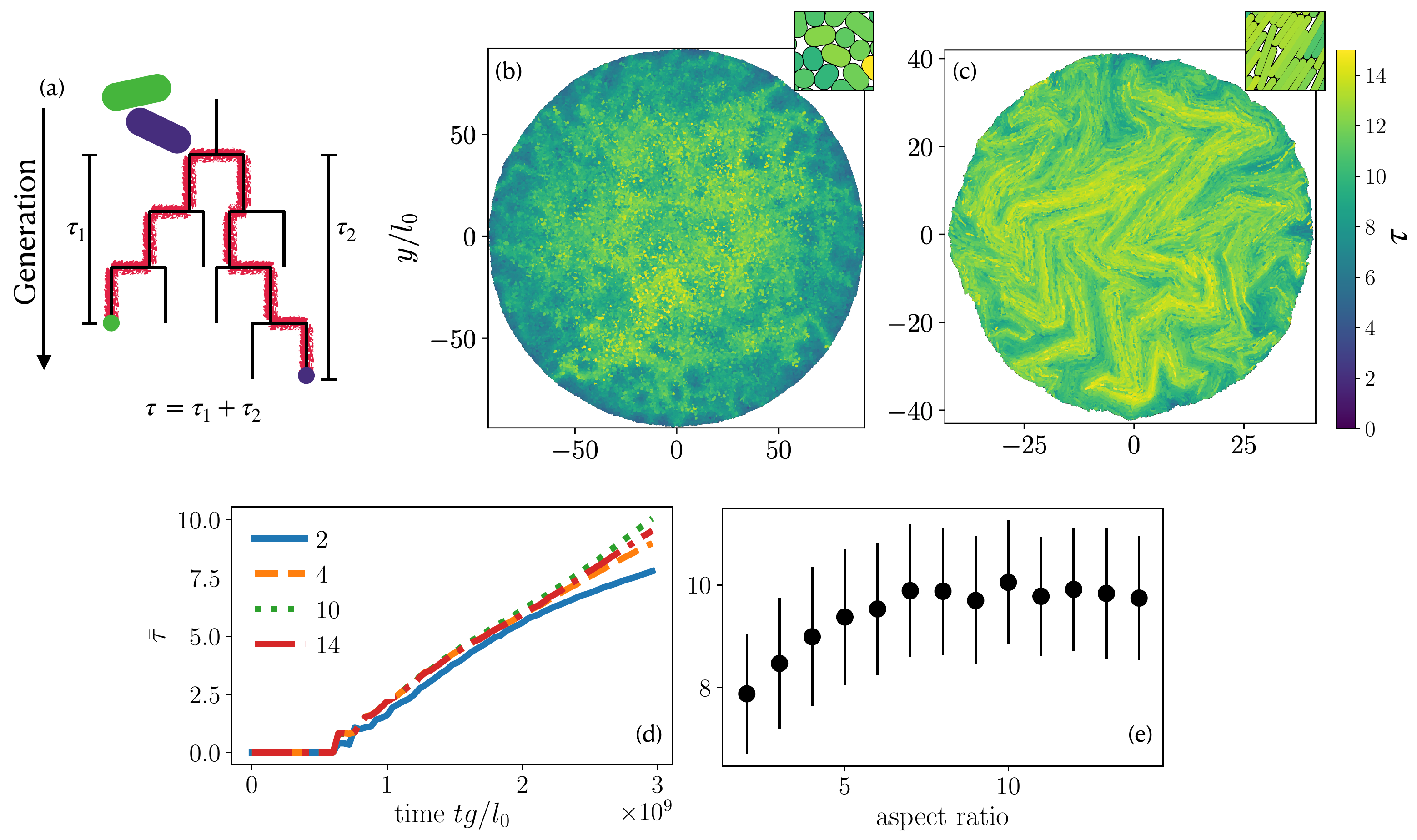}
    \caption{(a) {Schematic for calculation of phylogenetic distance $\tau$. The two particles (green and purple) are neighbors. Green and purple dots mark the particles of interest at present time, while red lines trace their ancestry back to the common ancestor.}
    (b-c) $\tau$ calculated for each cell in colonies with cell aspect ratio (b) $2$ and (c) $10$. 
    {Insets in (b) and (c) show a magnified view of the center of each colony.}
    (d) Colony average $\bar \tau $  of phylogenetic distance as a function of time  for particles of various aspect ratios as given in the legend. 
    (e) $\bar \tau$  at the end of our simulations for a range of aspect ratios. The error bars are the standard deviations of $\bar \tau$  in each colony.
    }
    \label{fig:mixing_colonies}
\end{figure}

We now turn to a population genetics perspective on mixing properties of our growing colonies, \textit{i.e.}\ the degree of relatedness for spatially nearby cells. In order to quantify mixing of particles, we compute 
$\tau$, the number of generations separating two cells phylogenetically, which is twice the number of generations since their most recent common ancestor
(Fig.~\ref{fig:mixing_colonies}a). Here, we first find the local neighbors of a {particle}
$i$ within a radius $r_{\mathrm{cut}}= \sqrt{\lambda d_0 l_0/\pi }$, where $\lambda=125$ controls the size of the local neighborhood. 
We note that the results presented in the following do not qualitatively change upon varying $\lambda$ within reasonable bounds. {For each particle $i$ we compute $\tau_i$ as the average of its $ \tau$ measure with each of its neighbors.}
In Fig.~\ref{fig:mixing_colonies}b,c we show $ \tau_i$ for colonies growing up to aspect ratios {$l_0/d_0$ of $2$ and $10$,} 
respectively. Visual inspection reveals that rods with a higher aspect ratio have a higher $ \tau_i$ and therefore mix more than at a lower aspect ratio.

On a global level, we determined the average phylogenetic distance $\bar \tau = \frac{1}{N}\sum_i \tau_i$. As expected, $\bar \tau$ increases with time (Fig.~\ref{fig:mixing_colonies}d). {Additionally, $\bar \tau$ increases with aspect ratio.} 
In the fully grown colony, we show $\bar \tau$ as a function of aspect ratio in Fig.~\ref{fig:mixing_colonies}e. 
We find that $\bar \tau$ increases for small aspect ratios, implying an increase of genetic mixing as aspect ratio increases, until around $l_0/d_0 = 7$, where $\bar \tau$ plateaus. 

\begin{figure}
    \centering
    \includegraphics[width=1.0\textwidth]{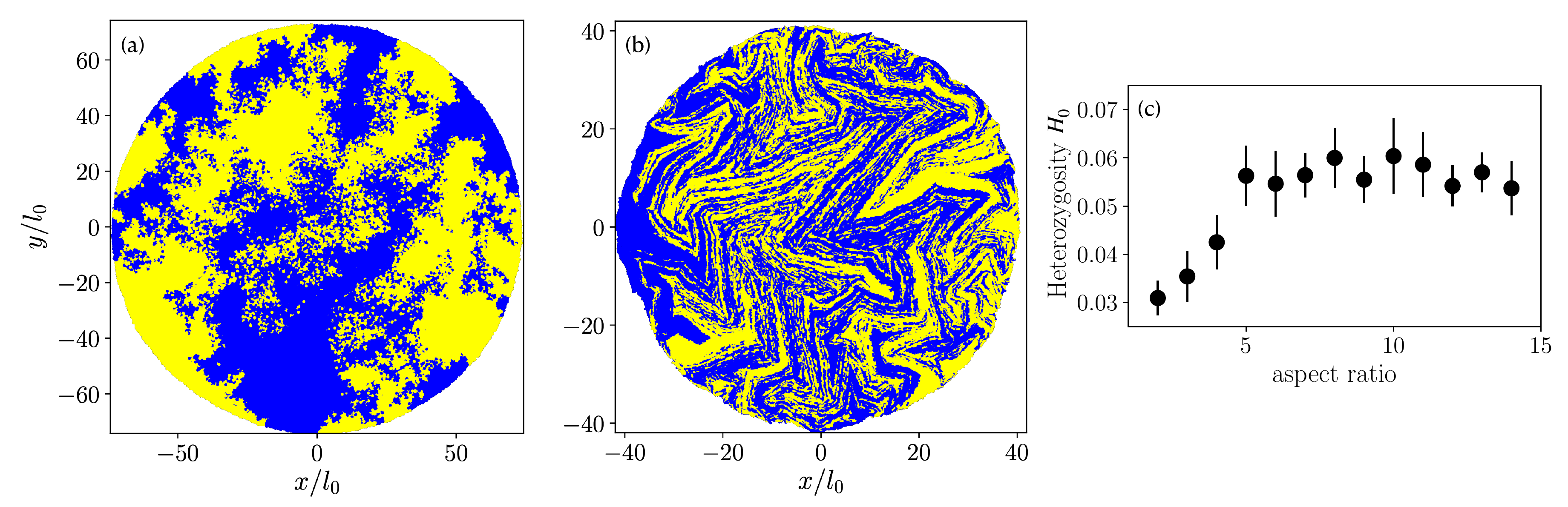}
    \caption{
    (a-b): Colonies with two alleles represented by yellow and blue colors, with cells having aspect ratio $2$ (a) or $10$ (b). 
     (c) Heterozygosity as a function of aspect ratio {at the end of each simulation}. 
    }
    \label{fig:mixing_obs}
\end{figure}

To make a connection to experiments~\cite{hallatschek2007genetic,farrell2017mechanical,korolev2010genetic} we performed simulations with two ``alleles'' (yellow and blue colors in Fig.~\ref{fig:mixing_obs}a,b) assigned randomly once the colony has reached $N=100$ particles. {Subsequently, cell division preserves the mother cell's allele in both daughter cells,} 
giving rise to the allele distributions shown in Fig.~\ref{fig:mixing_obs}a,b. As {with the $\bar \tau$ measure, } 
the alleles show that a colony with aspect ratio {10} 
(Fig.~\ref{fig:mixing_obs}b) is {better} 
mixed than a colony with aspect ratio {$2$.} 

{As a measure of genetic mixing in our two-allele representation, 
we compute the heterozygosity $H$, which, most generally, is the probability that two randomly selected organisms in the population share the same allele, but which can depend on distance $r$ in spatially structured populations \cite{korolev2010genetic}. $H$ is accessible to measurement in an experimental setting~\cite{hallatschek2007genetic,farrell2017mechanical,korolev2010genetic}. We study the small-distance limit of $H(r)$ averaged over all cells in the colony, which we call $H_0$, by considering all pairs of neighboring particles. With this restriction, the probability that two particles have different alleles equals the probability that both cells are on a line of contact between blue and yellow sectors. This recasts $H_0$ as a geometric measurement, $H_0= C_{BY}/A_{\mathrm{B}}$, where $C_{\mathrm{BY}}$ is the contour length of contact between blue and yellow sectors (including single-particle ``sectors''), and $A_{\mathrm{B}}$ is the total area occupied by the blue particles. 
Similarly to  $\bar \tau$, larger values of $H_0$ imply better mixing of the colony.
}

{For a given Brownian dynamics simulation, we measure $H_0$ for an ensemble of random blue/yellow color assignments when the colony has size $N=100$ particles, propagating the colors forward in time by following the simulation's phylogenetic tree, obtaining additional configurations similar to Fig.~\ref{fig:mixing_obs}a,b.
These are used to compute the colony-averaged value of $H_0$ and its associated standard deviation, shown in Fig.~\ref{fig:mixing_obs}c.
We find that for smaller aspect ratios, heterozygosity increases with increasing aspect ratio, 
implying an enhanced genetic mixing. 
Comparing our findings for $\bar \tau$ and $H_0$ to the measurement of nematic correlation length (Fig.~\ref{fig:snapshots}f), we see that at low aspect ratios, colony self-mixing improves with increasing aspect ratio just as nematic order begins to emerge.  
}


It is possible that simulations of larger systems would observe active nematic self-mixing more akin to the fixed-size case. The exponentially increasing computational expense with number of generations favors continuum models at these larger scales. Agent-based modeling such as the results we have presented can help to constrain and validate continuum hydrodynamic models for these purposes.

\section{Conclusions}

Motivated by the possibility that active mixing dynamics might counteract genetic demixing, we have presented results from particle-based computational modeling of growing colonies of immotile bacteria.
To the question of whether growing active nematics exhibit the self-mixing dynamics of fixed-size active nematics, a surprisingly ``mixed'' answer emerges from our findings. On one hand, growing active nematics bear many of the signatures of their fixed-size counterparts: not only do disclinations arise in $\pm 1/2$-winding number pairs, but the $+1/2$ defects behave like active quasiparticles that preferentially move along their ``comet-head'' direction, with an accompanying pair of counter-rotating vortices (for isolated defects) as predicted by hydrodynamic theory ~\cite{giomi2014defect}. On the other hand, the active self-propulsion of $+1/2$ defects is generally small compared to the colony's radial expansion flow, so that the trajectories of $+1/2$ and $-1/2$ defects appear very similar. Most importantly, $+1/2$ defect trajectories do not braid around each other; relatedly, we find no chaotic mixing in the flow field of bacterial cells. 


Despite the absence of chaos, we find, with measures inspired by population genetics, that active nematic colonies of high-aspect ratio cells mix themselves more efficiently than active isotropic colonies of rounder cells. Self-mixing quality improves with increasing aspect ratio up to around $\approx 7$, a trend seen both in the typical time since common ancestry $\bar \tau$ for spatially nearby cells as well as in the heterozygosity $H_0$, the probability for a cell to have a neighboring cell of different type. Because this physical self-mixing opposes the 
process of genetic demixing, our results are consistent with the hypothesis that active nematic dynamics {measurably alter evolutionary trends associated with colony growth for} 
bacterial species with elongated cell shapes. We therefore expect that loss of deleterious alleles and fixation (complete takeover) of advantageous alleles will occur more quickly in colonies of higher aspect ratio cells (all other parameters being equal) \cite{hallatschek2007genetic,korolev2010genetic}, a prospect that we intend to investigate in future studies.



The model presented here includes only the birth of new particles; in future work we aim to also include cell death and removal. For an equal birth and death rate, the model then becomes a typical active nematic (as in the simulations of Ref.~\cite{decamp2015orientational}). Additionally, the view of our system as a nematic fluid ignores orientational discontinuities along linear grain boundaries, which have been shown to play in important role in confined, two-dimensional smectic liquid crystals of hard rods~\cite{monderkamp2021topology}. Applying this perspective  and including grain boundaries into analysis of particle-based active nematics simulations might give new insights into the structure and growth of bacterial colonies. 

There is also a broad range of questions inspired by population genetics that can be probed with this model, which we have used here to examine only the simplest scenario of neutral evolution without mutation or competition for nutrients. Inspired by studies on lattice models and reaction-diffusion-type models \cite{edmonds2004mutations,klopfstein2006fate,korolev2010genetic}, other topics of interest to address with this model include: the fixation time of an advantageous allele; the survival time of a deleterious allele; the stability of a colony against ``mutational meltdown'' in the presence of both mutation and selection \cite{lavrentovich2013radial}; growth rate heterogeneity due to differential uptake of finite nutrient resources; and robustness of a genetically heterogeneous colony against a spatially or temporally heterogeneous environment \cite{mobius2015obstacles,beller2018evolution,ChuEvolution2019,gralka2019environmental}. {Agent-based simulations incorporating these genetic scenarios alongside mechanics are a promising tool for combining the perspectives of population evolution and active matter dynamics, providing insights into the evolutionary consequences of emergent collective motions.}



\section*{Conflict of Interest Statement}

The authors declare that the research was conducted in the absence of any commercial or financial relationships that could be construed as a potential conflict of interest.

\section*{Author Contributions}

F.J.S.\ and D.A.B.\ designed the work and wrote the paper. F.J.S.\ performed the work. 


\section*{Funding}
We gratefully acknowledge computing time on the Multi-Environment Computer for Exploration and Discovery (MERCED) cluster (NSF Grant No. ACI-1429783).

\section*{Acknowledgments}
We gratefully acknowledge helpful discussions with Jimmy Gonzalez Nu\~nez, Jordan Collignon, Mikhal Banwarth-Kuhn, and Suzanne Sindi.



\bibliographystyle{frontiersinHLTH&FPHY} 
\bibliography{bacteriabib}


\end{document}